\pdfoutput=1
\documentclass[english,manuscript]{aastex}
\usepackage[T1]{fontenc}
\usepackage[latin9]{inputenc}
\usepackage{graphicx}

\makeatletter
\shorttitle{The signature of LLAGNs in the nearby universe}
\shortauthors{Neri-Larios et al.}
\bibpunct[,]{(}{)}{;}{a}{,}{,}

\makeatother

\usepackage{babel}

\begin{document}

\title{The signature of LLAGNs in the nearby universe}

\author{D.~M. Neri-Larios, J.~P. Torres-Papaqui, R. Coziol, J.~M. Islas-Islas, and R.~A. Ortega-Minakata}

\affil{Departamento de Astronom\'ia, Universidad de Guanajuato, Apartado Postal 144, 36000 Guanajuato, Gto, M\'exico (daniel@astro.ugto.mx)}

% English abstract
\begin{abstract}
We have used the diagnostic diagram that compares the
ratio of emission lines [NII]$\lambda$6584/H$\alpha$ with the
equi\-va\-lent width of [NII]$\lambda$6584, as proposed by Coziol et
al. (1998), to determine the source of ionization of SDSS NELGs that
cannot be classified by standard diagnostic diagrams, because the
emission line [OIII]$\lambda$5007, H$\beta$, or both, are missing.
We find these galaxies to be consistent with low luminosity AGNs,
suggesting that this characteristic is the signature of the LLAGNs
in the nearby Universe.
\end{abstract}

% Keywords must be from the standard list and in alphabetical order.
% You should have no more than SIX different keywords.

% Spanish abstract - leave blank and it will be translated by the
% editors.
\section{Resumen}
Hemos utilizado el diagrama diagn\'ostico que compara la
raz\'on de las l\'ineas de emisi\'on [NII]$\lambda$6584/H$\alpha$
con el ancho equi\-va\-lente de [NII]$\lambda$6584, propuesto por
Coziol et al. (1998), para determinar la fuente de ionizaci\'on de
galaxias con l\'ineas de emisi\'on agostas tomadas del SDSS donde la
emisi\'on en [OIII]$\lambda$5007, en H$\beta$, o ambas, no est\'an
presentes y por esta raz\'on fueron excluidas de an\'alisis previos.
Estas galaxias se encuentran generalmente poblando la regi\'on del
digarama consistente con AGNs de baja luminosidad, sugiriendo que
esta caracter\'istica es la firma de los LLAGNs en el universo
cercano.

\section{Introduction and Discussion}
\label{sec:intro} The most common (standard) diagnostic diagrams
devised to identify the source of ionization of Narrow Emission Line
Galaxies (NELGs) in the nearby universe are those that compare the
line ratios [OIII]$\lambda$5007/H$\beta$ with
[NII]$\lambda$6584/H$\alpha$ [OI]$\lambda6300/H\alpha$, or
[SII]$\lambda\lambda6717,6731/H\alpha$. Applying different
separation sequences Kewley at al. (2001) and Kauffmann et al.
(2003) proposed to discriminate between NELGs ionized by thermal
photo-ionization, consistent with Star Forming Galaxies (SFGs), and
NELGs ionized by non thermal photo-ionization source, generally
called Active Galactic Nuclei (AGNs). Galaxies with intermediate
line ratios are classified as Transitory Objects (TOs).

The standard diagnostic diagrams are unfortunately useless in cases
where some of the emission lines listed above are missing. In the
SDSS survey this leaves an important fraction of galaxies
unclassifiable (Cid Fernandes et al. 2010). The main interest of
NELGs with emission lines missing is that they are particularly
frequent in dense galactic structures, like clusters of galaxies
(Phillips et al. 1986) and compact groups (Coziol et al. 1998;
Mart\'{\i}nez et al. 2010).

In Coziol et al. (1998) it was shown that even after subtracting
different templates from the spectra the missing lines do not
appear. This is why these authors have used an alternative
''diagnostic diagram'' to classify these galaxies. The NII diagram
compares the equivalent width (EW) of [NII]$\lambda$6584, which is
unaffected by the template subtraction, with the corrected by
template ratio [NII]$\lambda6584/H\alpha$.

From the SDSS DR5 we have downloaded a sample of 476931 NELGs
spectra, using the Virtual Observatory service\footnote{
http://www.starlight.ufsc.br}. Keeping only galaxies that have
emission lines S/N$\ge3$ (S/N$>$10 in the continuum) reduces our
sample to 224846 galaxies. In this last sample we count 34307
galaxies without H$\beta$, 12455 without [OIII] and 2840 without
both lines, which represents 22\% of the sample. These galaxies were
systematically discarded by Cid Fernandes et al. (2010) in their
recent study about the ``forgotten'' population of weak line
galaxies (WLGs). Note also that since our galaxies have S/N$\ge3$
they do not classify as WLGs.

\begin{figure}[!t]
  \includegraphics[width=0.9\columnwidth]{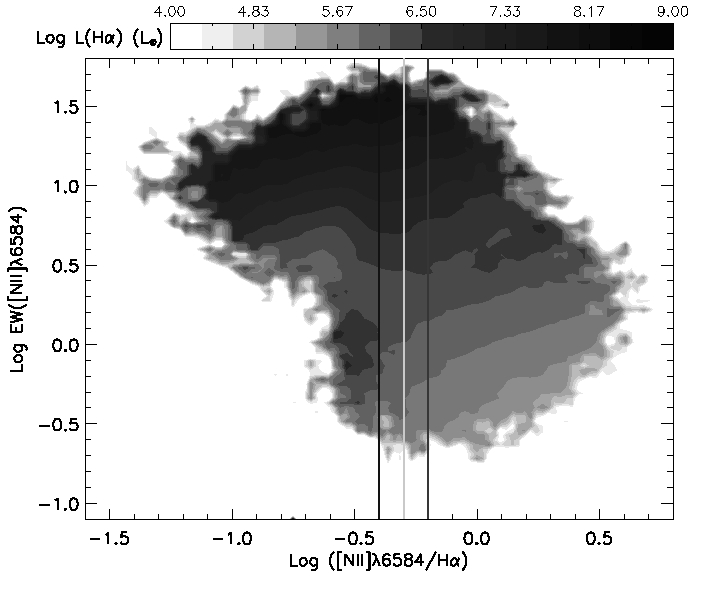}
  \caption{Contours of luminosity in H$\alpha$.
  The vertical line at log([NII]/H$\alpha$)=-0.3 separates SFGs from AGNs.
  The two lines at -0.4 and -0.1 define the buffer zone inhabited by TOs.}
  \label{fig:01}
\end{figure}

In Figure \ref{fig:01} we show the NII diagram for the NELGs with
the emission lines missing. 62.3\% of the galaxies without [OIII]
falls on the AGN side of the NII diagram (to the right of separation
line at -0.3 in [NII]/H$\alpha$). This fraction increases to 91\%
for the galaxies without both lines and 93\% for the galaxies
without H$\beta$. Therefore, most of the NELGs with emission lines
missing are AGNs. These results differ significantly from what Cid
Fernandes et al. (2010) have obtained. They found very few WLG are
AGNs. This suggests that the absence of emission lines points to a
different nature for the galaxies.

Most of the NELGs with emission lines missing have EW below
Log(EW[NII])$=0.6$: 77\% without [OIII], 94\% without H$\beta$ and
98\% without both lines. By definition, the EW is the ratio between
the flux in the emission line and the flux in the adjacent
continuum. This parameter is consequently sensitive to the
luminosity of the emission line and to the underlying stellar
population. A small value of EW suggests the galaxies have an
early-type morphology (see the article by Torres-Papaqui et al. in
this proceeding), and/or that the emission line has a low
luminosity. This is confirmed in Figure \ref{fig:01} where it is
seen that the H$\alpha$ luminosity decreases with the EW. The median
H$\alpha$ luminosity below Log(EW[NII])$=0.6$ is that of the
galaxies without H$\beta$, which is 5.6$\times$10$^{39}$ erg
s$^{-1}$. This property categorizes these galaxies as Low Luminosity
AGNs (LLAGNs).

\begin{figure}[!t]
  \includegraphics[width=0.8\columnwidth]{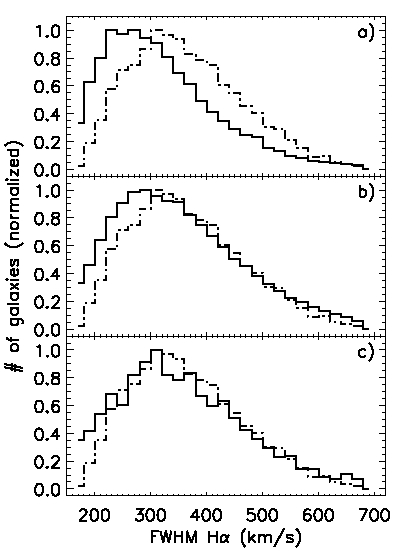}
  \caption{Distributions of the FWHM of H$\alpha$:
  a) NELGs without[OIII]; b) NELGs without H$\beta$;
  c) NELGs without both lines.
  The point-dashed line corresponds to the distribution of luminous AGNs.}
  \label{fig:02}
\end{figure}

Figure\ref{fig:02} shows that the NELGs with emission lines missing
have broad FWHM, similar to luminous AGNs. After correction for the
resolution of the instrument, the FWHM values fall between 170 up to
700 km s$^{-1}$. A higher fraction of the galaxies without [OIII]
have lower FWHM, which is consistent with the high number of TOs and
SFGs in this sample. The FWHM seems like another good criterion to
separate AGNs from SFGs.

We have found the NELGs with emission lines missing to be in
majority LLAGNs. Therefore, this characteristics--the absence of
emission lines--could be taken as the signature of LLAGNs in the
nearby universe (Coziol et al. 1998). Summing all the luminous AGNs
with the LLAGNs and TOs in our sample suggests that almost half of
the NELGs in the nearby universe are AGNs (Miller et al. 2003). This
result seems now much more consistent with the high number of QSOs
observed at high redshifts. This observation may support the
standard interpretation of AGNs (independent of their luminosity) as
galaxies with active black holes in their nucleus.

We also acknowledge support from PROMEP (Grant No. 103.5-10-4684).

\end{document}